# Analysis of an Extension Dynamic Name Service
# A discussion on DNS compliance with RFC 6891

Ivica Stipovic
Ward Solutions
Dublin, Republic of Ireland
e-mail: ivica.stipovic@ward.ie

*Abstract*—Domain Name Service (DNS) resolution is a mechanism that resolves the symbolic names of networked devices to their corresponding Internet Protocol (IP) address. With the emergence of the document that describes an extension to a DNS service definition, it was becoming apparent that DNS implementations will need to comply with some modified DNS behaviour. One such modification is that the DNS continues to use the User Datagram Protocol (UDP) to transmit DNS payloads that are longer than 512 bytes. Until the emergence of the Extension DNS (EDNS) specification, DNS servers would switch over from UDP to Transmission Control Protocol (TCP) if the response payload was larger than 512 bytes. With the new EDNS capability, it was required that DNS replies would continue to provide responses as UDP datagrams even though the response was larger than 512 bytes. To the author's best knowledge, there are no academic articles dealing with the assessment of the DNS servers against EDNS specification. This paper examines the level of compatibility for a number of public DNS servers for some popular internet domains. It also explores behaviour of some contemporary DNS implementations such as Microsoft Windows 2012, 2016 and 2019 as well as Linux-based BIND in regards to the EDNS.

## I. INTRODUCTION

Domain Name System (DNS) is a system responsible for the device name resolution that translates symbolic names of devices into their corresponding IP address. These names are often referred to as Fully Qualified Domain Names (FQDN) as they consist of the name of the specific device, usually appended with their belonging domain.

DNS was developed to help humans navigate through media such as the internet where remembering the IP address would be too counter-intuitive and difficult. From the early days, DNS specification was devised to help software vendors develop and implement standardised operations aimed at resolving hostnames to their corresponding IP addresses. It was essential that various DNS implementations be capable of interoperating with each other to provide a seamless flow of name resolution across different authoritative domains. DNS is organized as a hierarchical structure with different owners across different domains.

This is why a Request For Comment (RFC) 1035 [1] was developed as an initial step to help standardise the DNS behaviour. Among other provisions, RFC 1035 defined that UDP messages should not exceed 512 bytes. Over time, DNS has evolved and introduced advanced features in terms of additional types of records such as TXT and DNSSEC. These two record types were not existent in the time when RFC 1035 was defined. TXT and DNSSEC are examples of records which require the DNS response payload size that often times exceeds 512 bytes. TXT is a type of DNS record that contains an arbitrary string and is defined in the RFC 1464 [3].

DNSSEC is another DNS record type designed with security in mind. Its primary purpose is to allow verification that the domain that sourced certain internet traffic is indeed a genuine domain. The relevance of DNSSEC record is typical for e-mail servers that may check if the source domain of an email is not a spoofed or malicious domain under the control of an attacker. This makes the DNSSEC record a good means to counter-fight email spamming. DNSSEC was defined in the RFC 4033 [4]. At the same time, TXT and DNSSEC have become regular queries from DNS clients to DNS servers or between DNS servers themselves.

There are some DNS operations such as Zone Transfer which require transmission over TCP. The reason why DNS Zone transfers use TCP protocol is that they transmit a lot more information than a simple UDP query. Zone transfer will usually transmit the content of the whole domain zone or a part of the zone. This means multiple records will be transmitted. TCP transmission is more reliable than UDP because TCP is a more reliable protocol which has an inherent mechanism to recover corrupted or lost network packets. UDP would need to rely on the application protocol to handle these transmission errors. Zone transfers occur between DNS servers of the same domain mostly to ensure that there is a fault tolerance in case a primary DNS server fails. The main objective of Zone Transfer is to ensure that all the zone records are transmitted to another designated DNS server. Even though it is possible to initiate zone transfer from an end-user DNS client, it is not considered a part of regular functionality that should be allowed. This is why many corporate policies will deny zone transfers initiated from end-user DNS clients and allow it only to and from the designated DNS servers. The new standard emerged to describe the details of how DNS should handle the packets whose size exceeded 512 bytes in the form of RFC 6891 [2]. It was labelled as an Extension Mechanisms for DNS (EDNS(0)).

## II. LITERATURE REVIEW

Besides the RFC 6891 definition, there is little other research available that investigates the compliance of DNS servers against EDNS feature. One of the online tools that allows easy verification of your DNS server is the one that Internet System Consortium provides called EDNS compliance checker [5]. The main deficiency of the online tool is that you cannot use it to check your internal DNS servers, unless they are exposed to the internet.

Another valuable resource that deals explicitly with EDNS issue is the "DNS Flag Day" campaign [6]. This campaign actively supports the alignment of public DNS servers with a number of standard behaviours, one of them includes EDNS compatibility. DNS Flag day also maintain the repository on Github [7] with a number of code snippets. Internet System Consortium also maintains BIND DNS server [8]. BIND was one of the DNS server implementations whose EDNS compatibility was examined in this research. Microsoft also maintains their website dedicated specifically to EDNS compatibility [9]. The research had to include DNS client applications too, as both UDP and TCP sessions will be dependent on the specific DNS client capabilities. Prior to the testing, some initial research was aimed at the two most popular DNS client applications – "dig" [10] and "nslookup" [11], respectively.

The research indicated some clear disadvantages of nslookup compared to dig [12][13][14]. There is little research available in terms of academic work that deals with the specific evaluation of DNS behaviour when UDP transmission deals with packets bigger than 512 bytes.

## III. METHODOLOGY

The methodology designed for the EDNS compatibility assessment had to take into account several factors that were hypothesised to contribute to the overall DNS behaviour.

One of the basic requirements was to test various DNS server software packages. To address this, the assessment included both Windows and Linux based DNS implementation, Microsoft DNS for Windows 2012, 2016 and 2019 and Linux BIND 9, respectively. While this was easy to achieve in testing the locally controlled servers, it was much more challenging to find and identify different flavours of DNS server on the internet. While there are tools that allow fingerprinting and version identification of DNS servers, many of the public DNS servers will be operating behind firewalls or other traffic filtering devices that prevent reliable version detection.

To address the challenge of covering multiple DNS software packages on the internet, the decision was made to run EDNS checks across several different internet domains. The assumption taken was that with multiple different internet domains, the results would be processed with different DNS server implementations.

Another level of inspection was introduced that included both implementations of DNS servers in the local test environment and public DNS servers on the internet. The reason for testing local and internet DNS servers was an opportunity to compare and correlate results for both domains. This inspection would facilitate the identification of potential discrepancies in configuration between these two groups of servers. Since the DNS resolution process involves both DNS client application that sources the query and DNS servers that reply to query, it was necessary to inspect at least two different implementations of DNS clients.

To facilitate multiple assessment criteria, two DNS client were used – "dig" [10] and "nslookup"[11]. While the "dig" client is implemented predominantly on Linux and UNIX operating systems, nslookup is equally present in both Windows and Linux-like operating systems. This is why "nslookup" was run from both Linux and Windows operating systems. Using two different DNS clients on two different operating systems allowed for potential detection of differences in behaviour that stems from the different operating system implementations. One more tool used to check the consistency of the results generated by nslookup and dig was the EDNS compliance checking engine run by isc.org and hosted on their website [5]. Even though the output of this automated checker was valuable, its continuous lack of availability during the test period excluded it from the analysis of the results.

## IV. RESULTS

The results of the EDNS compliance are summarised in the Table 1 shown below.

| Domain | Number of DNS servers per domain | EDNS(0) compliance nslookup | EDNS(0) compliance dig |
|--------|----------------------------------|-----------------------------|------------------------|
| Microsoft.com | 5 | No | No |
| Redhat.com | 6 | No | Yes |
| Oracle.com | 6 | No | Yes |
| Verizon.com | 5 | No | Yes |
| Isc.org | 4 | No | Yes |
| Local Windows 2012 | 1 | No | Yes |
| Local Windows 2016 | 1 | No | Yes |
| Local Windows 2019 | 1 | No | Yes |
| Local BIND 9 Linux | 1 | No | Yes |

Table 1 EDNS compliance of some DNS implementations

The first column labelled "Domain" indicates the internet domain that was inspected. The column "Number of DNS servers per domain" indicates the number of DNS servers that were found authoritative for this particular domain.

This number is not equal for every domain as a different number of secondary DNS servers can be defined, as the domain owner deems appropriate. The number of DNS servers per domain will be equal to the number of Name Server (NS) records defined per domain. The last four rows show locally installed Windows DNS and Linux DNS servers. Having locally installed DNS server is the reason why there is only one server per domain. All queries run via dig in the test were against TXT and DNSSEC. Nslookup queried only for TXT records. The reason for this difference is that nslookup does not support queries against DNSSEC records. The main event in the DNS session that determined the switchover from UDP to TCP transmission was observed which sets the flag in the application layer indicating the message is truncated. This is shown in Figure 1 below.

```
User Datagram Protocol, Src Port: 53, Dst Port: 59638
Domain Name System (response)
    Transaction ID: 0x0004
  Flags: 0x8780 Standard query response, No error
    1... .... .... .... = Response: Message is a response
    .000 0... .... .... = Opcode: Standard query (0)
    .... .1.. .... .... = Authoritative: Server is an authority for domain
    .... ..1. .... .... = Truncated: Message is truncated
    .... ...1 .... .... = Recursion desired: Do query recursively
    .... .... 1... .... = Recursion available: Server can do recursive queries
```
Figure 1 Truncated bit set in the DNS payload

This truncated bit will be set in the UDP packet and by inspecting the session logic in the network sniffer, one can see that the session continues with establishing the TCP connection between the client and the server. Note that the DNS client will automatically initiate TCP handshake connection with DNS server after it detects the truncation. This TCP switchover is shown in Figure 2 below.

```
Destination     Protocol  Length  Info
192.168.56.103  DNS       78      Standard query 0x000b TXT test.local.ward.ie
192.168.56.103  DNS       70      Standard query 0x000c TXT test.local
192.168.56.1    DNS       131     Standard query response 0x000c TXT test.local
192.168.56.103  TCP       66      3699 → 53 [SYN] Seq=0 Win=64240 Len=0 MSS=146
192.168.56.1    TCP       66      53 → 3699 [SYN, ACK] Seq=0 Ack=1 Win=8192 Len
192.168.56.103  TCP       54      3699 → 53 [ACK] Seq=1 Ack=1 Win=525568 Len=0
```
Figure 2 TCP handshake initiation after the Truncated bit

Note the "SYN" flag in the initial TCP handshake sent to destination port 53 (DNS).

Most online servers' responses confirmed they are EDNS compliant but only if dig is used as a client application. The only exception here is Microsoft DNS servers which failed to comply to EDNS irrespective of the DNS client used. The indication of the truncated DNS message is displayed in the output from the "dig" in Figure 2 below.

Figure 3 output from dig client during the assessment of a public DNS indicating non-compliance with EDNS

All offline lab servers' responses were EDNS compliant, but only if dig was used as a DNS client. Nslookup returned the non-compliant responses for all the servers, including local and online. The EDNS checker was not available all the time during the assessment despite its valuable results and is therefore not included in the analysis of the results.

V. SECURITY IMPLICATION OF NON-COMPLIANCE

One potential issue of EDNS non-compliance is security related. Corporate security and firewall policies may dictate that only the UDP protocol is allowed for end-user DNS queries. End-user queries assume only the DNS queries sourced by end-user applications such as internet browsers, email clients, dig and nslookup. Even if the end-user queries for TXT or DNSSEC records that can exceed 512 bytes, the DNS response should still come via UDP protocol if the destination server is EDNS compliant. Therefore, the corporate firewalls may be configured to reject any DNS communication over TCP, except maybe between the primary and secondary authoritative DNS servers. If the end-user client uses nslookup or the destination DNS server is not EDNS compliant and strict firewall rule is in place, the query will fail if the response payload exceeds 512 bytes. Such a firewall rule can then introduce an inadvertent denial of service where the name resolution process may fail. .

VI. RISK RATING

The security implications of this non-EDNS compliant behaviour identified so far are related to the denial of service. The denial of service due to a firewall rule that does not permit TCP DNS will cause the end-user DNS queries to fail. The possible mitigation could allow the TCP transmission of DNS queries. While such a rule may relax somewhat the security posture of the company, no other attack vectors have been identified. One possible compensating measure for allowing DNS via TCP can be an IP based restriction configured on the DNS servers. This IP based restriction will allow the zone transfers only from predefined IP addresses. These predefined IP addresses will belong to the company's approved and trusted DNS servers only.

The risk rating proposed for this EDNS non-compliance is Medium to Low.

## VII. CONCLUSION

This paper demonstrated that even though the Extension DNS (EDNS(0)) mechanism has been established as the mandatory requirement for DNS implementations, there are some caveats causing the lack of compliance. One of the causes is using the nslookup DNS client. Nslookup is nowadays an obsolete DNS client which lacks some important capabilities, but is still widely used. Another cause for EDNS non-compliance is that some public DNS servers do not support EDNS(0) as demonstrated in the results section. Any discrepancy in implementing EDNS(0) compatibility may cause a denial of service due to the firewall rules that assume the compliance whereas some scenarios clearly demonstrate this is not always the case.